\documentclass{aa}
\usepackage{graphicx}
\usepackage{txfonts}

\usepackage{graphicx}
\usepackage{natbib}
\usepackage[colorlinks = true,linkcolor = blue,urlcolor = blue,citecolor = blue,bookmarks = true,breaklinks]{hyperref}

\begin{document} 

\title{Balmer continuum  enhancement detected in a mini flare observed with IRIS}
\author{Reetika Joshi
\inst{1,2}\and
Brigitte Schmieder
\inst{1,3}\and
Petr Heinzel
\inst{4}\and
James Tomin
\inst{1}\and
Ramesh Chandra 
\inst{2}\and
Nicole Vilmer
\inst{1}
}
\institute{
LESIA, Observatoire de Paris, Université PSL, CNRS, Sorbonne Université, Université de Paris, 5 place Jules Janssen, 92190 Meudon, France\\
\email{reetikajoshi.ntl@gmail.com}
\and
Department of Physics, DSB Campus, Kumaun University, Nainital -- 263 001, India
\and
Centre for mathematical Plasma Astrophysics, Dept. of Mathematics, KU Leuven, 3001 Leuven, Belgium
 \and
Astronomical Institute of the Czech Academy of Sciences, Fri\v{c}ova 298, 251 65 Ond\v{r}ejov, Czech Republic
}

 %
\abstract
{{Optical and near-UV continuum} emissions  
in  
flares contribute substantially  to   flare energy budget. 
  {Two mechanisms play an important role for continuum emission in flares: hydrogen recombination after sudden ionization at chromospheric layers  and  transportation of the energy radiatively from { the chromosphere} to lower layers in the atmosphere, the so called  {\it back-warming}.}}
  {{The aim of the paper is to disentangle  between these two mechanisms  for  the excess of Balmer continuum  observed in a flare.}}
   %
 {We combine the observations of   Balmer continuum obtained with  the Interface Region Imaging Spectrograph  (IRIS)   (spectra and slit-jaw images (SJIs) 2832 \AA) and   hard X-ray (HXR) emission detected by  FERMI  Gamma Burst Monitor (GBM) during a mini flare. Calibrated Balmer continuum is compared to non-LTE radiative transfer flare  models  and radiated energy is estimated.  {Assuming thick target HXR emission}, we calculate the energy of non-thermal electrons  detected by  FERMI GBM  and compare it to  the radiated energy.}
   {The favorable argument of a relationship between the Balmer continuum excess and the HXR emission is that there is a good time coincidence between both of them. In addition,  the shape of the maximum brightness in the 2832 SJIs, which is mainly due to this Balmer continuum excess, is similar to the FERMI/GBM light curve.
   The electron-beam flux estimated from FERMI/GBM  between 10$^9$ to 10$^{10}$ erg s$^{-1}$ cm$^{-2}$ is consistent with the beam flux  required in {non-LTE radiative transfer  models}  to get the  excess of Balmer continuum  emission observed in this IRIS spectra.}
   {The low energy input by non thermal electrons  above 20 keV {is  
   sufficient} to  produce  the enhancement of Balmer continuum emission. This  could be explained by 
   the topology of the reconnection site. 
   {The reconnection starts in a tiny bald patch region which is transformed dynamically in a X-point current sheet. The size of the interacting region
   would be under the spatial resolution of the instrument.}}
   
   \authorrunning{Reetika Joshi et al.} 
   \titlerunning{Balmer continuum enhancement at the base of a solar jet}
   \maketitle
%
\section{Introduction}\label{sec:intro}
Heating of the lower solar atmosphere during solar flares is an interesting and still open area in solar Physics, as it deals with the energy distribution in  flares. That is {directly related  to the radiation emitted during a solar flare.
The heated atmosphere produces the enhancement of emission in many lines and continua.
The origin of  optical continuum  in flares comes from two mechanisms: the hydrogen recombination as for continua (Paschen, Balmer) in the chromosphere   and H${^-}$ emission in the photosphere.} The optical continuum enhancement 
is a reliable signature of the  so-called white light flare \citep[WLF,][]{Fletcher2011}.
Enhancement of the Balmer continuum below 3646 \AA\
has been reported in ground based observations of flares \citep{Neidig1989} and recently 
in the  Interface Region Imaging Spectrograph \citep[IRIS,][]{Pontieu2014} spectra during  strong X-class WLF 
\citep{Heinzel2014,Kleint2016, Kleint2017}.  Electron beams are  often invoked to explain the  WLF heating 
in the {low chromosphere} and photosphere. The enhancement of the Balmer continuum emission is produced in higher levels 
in the chromosphere.  
In \citet{Heinzel2014} there is a comparison of the enhancement of the Balmer continuum with the light curves from RHESSI
and a good spatial and temporal correlation is found, which is explained by electron-beam heating
and ionizing  the chromosphere. Then the subsequent recombination leads
to Balmer-continuum emission. It is the chromosphere which could produce the Balmer continuum emission enhancement, while the photosphere may in some cases
produce the white-light emission which is probably due to {radiative} backwarming \citep{Machado1989}. This issue was discussed {earlier in \citet{Ding2003, Heinzel2014} and quantitatively modelled in \citet{Kleint2016}.}

\begin{figure*}[ht!]
\centering
\includegraphics
[width=\textwidth]
{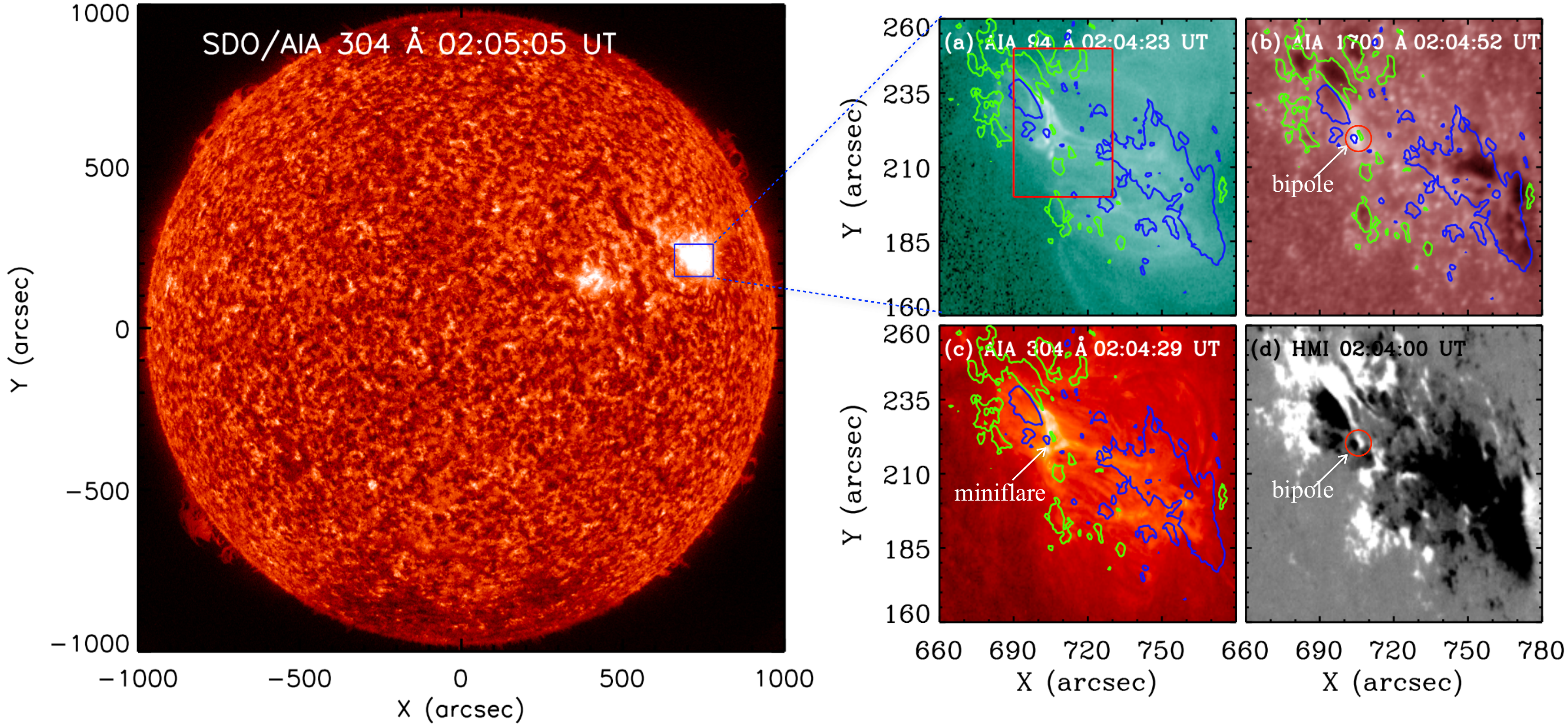}
\caption{Active region AR NOAA 12736  observed with AIA on March 22, 2019. ({\it Left panel}): full disk in 304 \AA, the {blue} box (field of view of the right panels) contains  the only AR  visible on the disk on that day.  ({\it Right panels} a, b, c):  images of the AR in different AIA filters, respectively in 94 \AA, {1700 \AA\,} and  304 \AA\  superimposed with the contours of  the magnetic field {of strength $\pm$ 300 Gauss}. Panel (d):  magnetic field observed  with HMI. The bipole  (small circle in  panels  (b) and (d))  is responsible of the reconnection.  The  mini flare  is indicated  by an arrow in panel (c).  The box in panel (a) is the field of view of Fig. \ref{fig:iris}.}
\label{fig:fulldisk}
\end{figure*}
A quasi-continuous enhancement in flares can be interpreted in three different ways. (i) it may be the Balmer continuum emission, superposed over the background spectrum. (ii) the continuum is due to contribution from the photosphere either by direct beam bombardement (but this could take place only for strong beams, such as in strong flares) or by radiative backwarming. (iii) finally the enhancement could be influenced by the emission of broad wings of Mg II lines. \citet{Heinzel2014} {took the} advantage to  use the  IRIS  2832 \AA\  band  spectra where the {Mg II wings were}
not present to {identify the real Balmer continuum excess at an  X-class flare site.}

{Recently a micro flare  observed in multiple wavelengths by SDO/AIA and IRIS  puzzled us as the Balmer continuum seems also to be enhanced, even in a  weak flare. 
{This micro  flare {that we called mini flare in the previous papers} occurred at the base of a solar jet on March 22, 2019. 
The analysis of the vector magnetic field revealed the existence of a flux rope in the vicinity of the jet and the  magnetic  reconnection took place in a bald patch 
region where  magnetic field lines were {tangential} to the photosphere \citep{Joshi2020FR}}. 

The proper motions of the photospheric magnetic  polarities suggested that  the twist of the flux rope was transferred during reconnection to the jet generating a twisted jet. 
 A detail spectroscopic analysis of the mini flare associated to this jet was carried out with  IRIS spectra  \citep{Joshi2021},  Using  the cloud model technique   applied to the  Mg II line profiles   we identified explosive clouds with supersonic Alfv\'en velocities.  
 {A stratification thermal model of the atmosphere at the time of the reconnection  was  proposed.}
The Mg II spectra show extended 
wings like  in IRIS bombs \citep{Peter2014,Grubecka2016}.  Twist and  bald patch reconnection  was confirmed with the analysis  of the  IRIS spectra (Si IV, C II, Mg II) by \citet{Joshi2020FR}.}

In this paper we {extend the IRIS  data analysis}  of the mini  flare {studied in \citet{Joshi2020FR, Joshi2021}} and present the Balmer continuum  around 2832  \AA\ observed with IRIS 
(Sect. \ref{sec:obs}). Section \ref{sec:iris} presents the IRIS  spectra and {2832 \AA\ slit-jaw images (SJIs)} analysis.
{Further, we analysed the {hard X-ray (HXR) emission 
{detected by} FERMI/GBM} (Sect. \ref{sec:fermi})}. 
We discuss on the possible {non-LTE} radiative transfer models, proposed in \citet{Kleint2016}, {deriving the radiated energy  from the measurements of the Balmer continuum.
We compare the  radiative energy to 
{the energy contained in non-thermal electrons as derived from HXR measurements}.} 
In conclusion  we conjecture that the Balmer continuum enhancement is
due to the {
energy electron}  deposit produced during the flare (Sect. \ref{dis}). {We suggest that 
the energy deposit, even with a relatively low value {is 
sufficient} because the reconnection is located in the low layers of the atmosphere in a {tiny} bald patch region} as  the magnetic topology of the region revealed \citep{Joshi2020FR}.
\begin{figure*}[ht!]
\centering
\includegraphics[width=\textwidth]{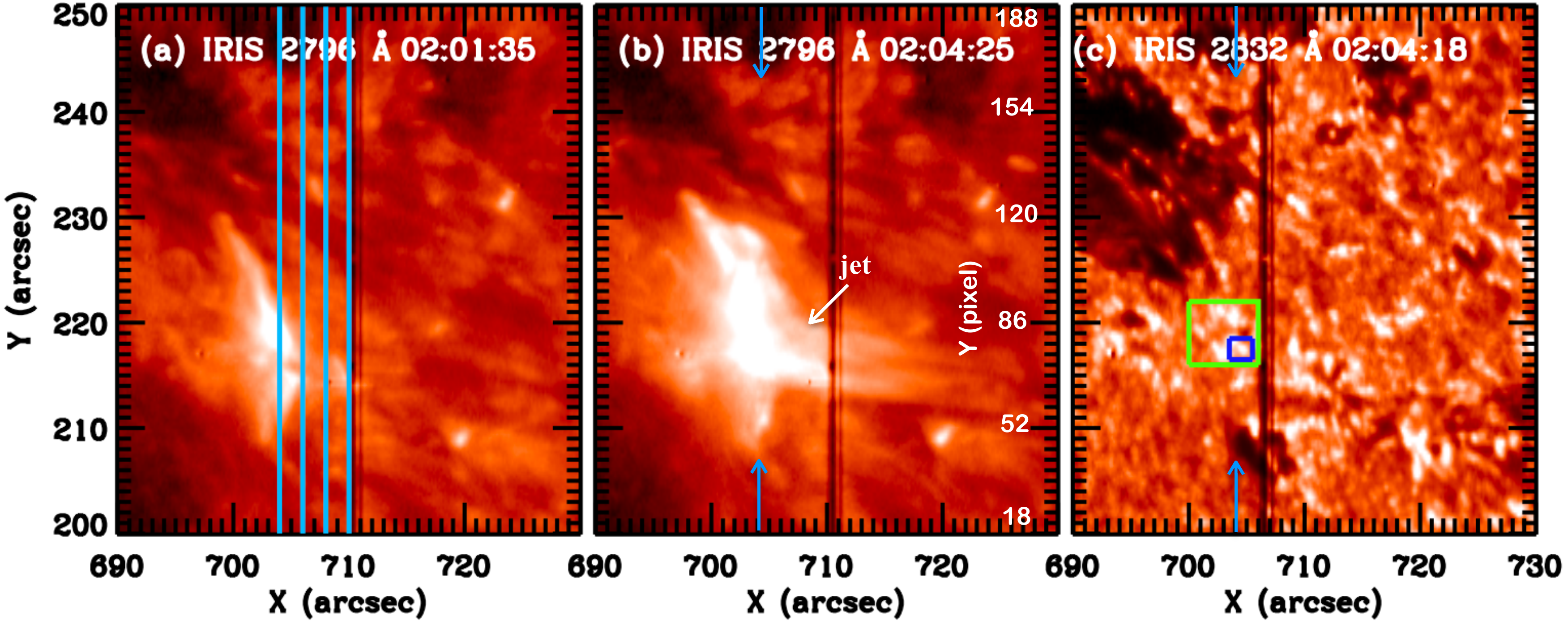}
\caption{{IRIS SJI of the mini flare 
in  2796 \AA\  (panels a-b) and 2832 \AA \  (panel c) filters respectively. In panel (a)  the four vertical cyan lines  are the four IRIS slit  positions of  the rasters}. {The vertical cyan arrows in panel (b-c) 
indicate the IRIS slit position 1, which crosses the mini flare (bipole) region presented in Fig. \ref{fig:fulldisk} (panels b- d). The green and blue boxes in panels (c) are the fields of view used to compute the light curves of the bright point} in Fig. \ref{goes}.
}
\label{fig:iris}
\end{figure*}
\begin{figure*}[ht!]
\centering
\includegraphics[width=\textwidth]{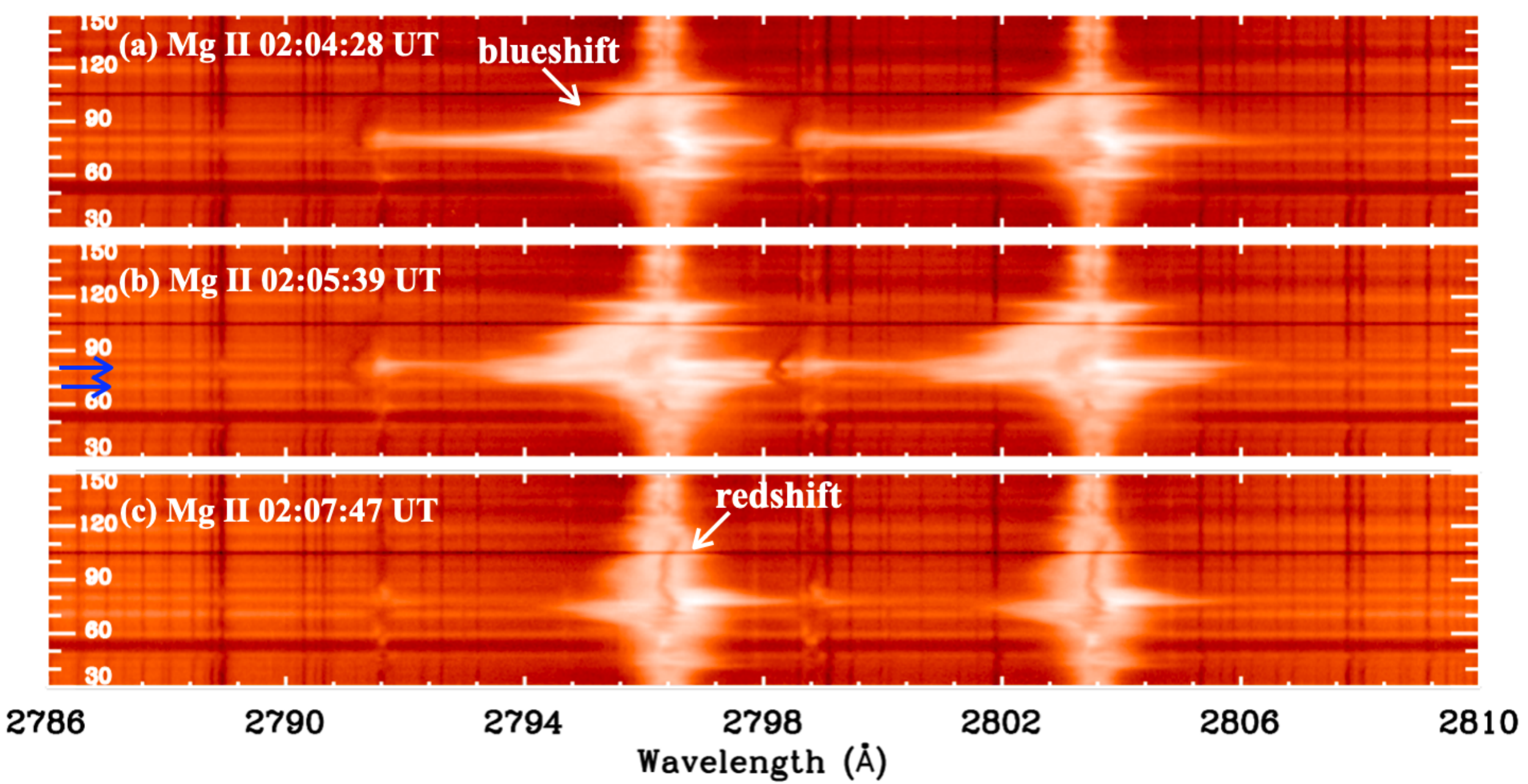}
\caption{Spectra of Mg II lines h and k in the Balmer continuum  full wavelength range of IRIS (50 \AA) during and after the mini flare  times at the reconnection site    along the first slit  position (Fig. \ref{fig:iris} panels a-c)). {Bidirectional outflows} are observed at the reconnection point (y=80 in panel a).  Balmer continuum  emission  is  visible in the three panels  at  two different pixels (y= 70, 80)  in wavelengths  10 \AA\ far from  the k line  core (see in panel b the  two  horizontal blue arrows).   
  Blueshifts, and redshifts  indicated by  two  white   arrows in  the Mg II k  line  wings correspond to  the  twisted  expelled jet  and   material falling back (panels a and  c).}
\label{balmer_MgII}
\end{figure*}
\section{Co-temporal observations} \label{sec:obs}
\subsection{AIA and IRIS SJI}
The mini flare (GOES B6.7) occurred  in the active region  (AR NOAA 12736), which was located at N09 W60 on March 22, 2019. It  was the only AR in the whole solar disk on that day
(Figure  \ref{fig:fulldisk} left panel). 
The {\it Atmospheric Imaging Assembly} \citep[AIA,] [] {Lemen2012} on board  the \emph{Solar Dynamics Observatory} \citep[SDO,] [] {Pesnell2012} allows us to display the images of the mini flare in  all the channels  
from UV (1600 and 1700 \AA) to EUV (94 \AA, 131 \AA, 171 \AA, 193 \AA, 211 \AA, and 304 \AA) covering a broad temperature range from  4500 K to 10$^6$  K
(Fig. \ref{fig:fulldisk} a-c).  The  1700 \AA\ filter   contains  hotter coronal lines which  are  in emission in strong flares  leading  to enhancement of  flare  brightenings  in this filter
\citep{Simoes2019}. Our mini flare being  a   weak flare  we may guess for  a  high probability {that the enhancement visible at 1700 \AA\ is due to heating of the plasma in the minimum temperature region around 4500 K.
The  topological configuration of the  AR  was analysed previously \citep{Joshi2020FR} showing that the AR formed by  successive emerging magnetic fluxes  (EMFs)  adjacent to each other.  The inversion line between two opposite polarities  belonging to two different EMFs was   the site of strong shear and    magnetic reconnection,  a  small bipole was identified as the origin of the mini flare  in  HMI magnetograms (Fig. \ref{fig:fulldisk} d). 
The mini flare   is in the central part of a bright, more or less north-south  semi circle overlying the inversion line between positive and negative polarities.}\\
IRIS contains three CCDs for the far ultra-violet (FUV) and near ultra-violet (NUV) {spectra} and one CCD for the SJIs.
{IRIS performed medium  coarse rasters  of 4 steps from 01:43:27 UT to
02:42:30 UT  centered at x=709$\arcsec$ and y=228$\arcsec$ in the  {SJI  field of view of
60$\arcsec$ $\times$ 68$\arcsec$.
The raster
step size is 
2$^{\prime}$$^{\prime}$ so each spectral raster spans a field of view
of 6 $^{\prime}$$^{\prime} \times$ 62
$^{\prime}$$^{\prime}$.  The nominal spatial resolution is 0.$^{\prime}$$^{\prime}$33.}
}   
IRIS SJIs 
in Mg II 2896 filter and 
in 2832 \AA\ continuum  filter  are recorded with a  14 sec cadence  (see  examples in Fig. \ref{fig:iris} a-c). The  four slits are drawn in   panel a. The first slit on the left crosses the mini flare. Slits 2, 3, 4 cross the jet on the right. In  2832 continuum  filter small bright dots are observed at the location of the mini flare {(green and blue boxes).} For an accurate comparison between the AIA  images and IRIS SJIs we  manually align these images by shifting the IRIS SJIs by 4$\arcsec$ in x-axis and 3$\arcsec$ in y-axis, {as  noted in the previous paper \citep{Joshi2021}}.
{The relationship  between heliographic coordinates and pixels along the slit of the spectra is shown in Figure \ref{fig:iris} (b).}

The IRIS Mg II  spectra during the  mini flare  
were analysed in \citet{Joshi2021}.  {Bidirectional flows} ($\pm$ 200 km s$^{-1}$) were observed at the flare reconnection site covering one or two pixels y= 79-80  at 02:03:46 UT.  At 02:04:28 UT strong  blueshifts  in pixels $> 80$ and redshifts a few minutes later {were also observed}, which  we interpret as material of the jet going up and then falling back.  The mini flare is definitively not a double-ribbon flare with many ribbons as it is  the case in \citet{Kleint2017}. A bright  continuum  is detected  all along the wavelength domain  in the Mg II spectra at  the  flare site {(y = 70 and 80 pixel) during a few minutes  (Fig. \ref{balmer_MgII}).}  In this study we focus on the Balmer continuum far away from the extended Mg II line wings  until  2835 \AA\ to be free of the emission of the Mg II line wings.

\begin{figure}[t!]
\centering
\includegraphics[width=0.5\textwidth]{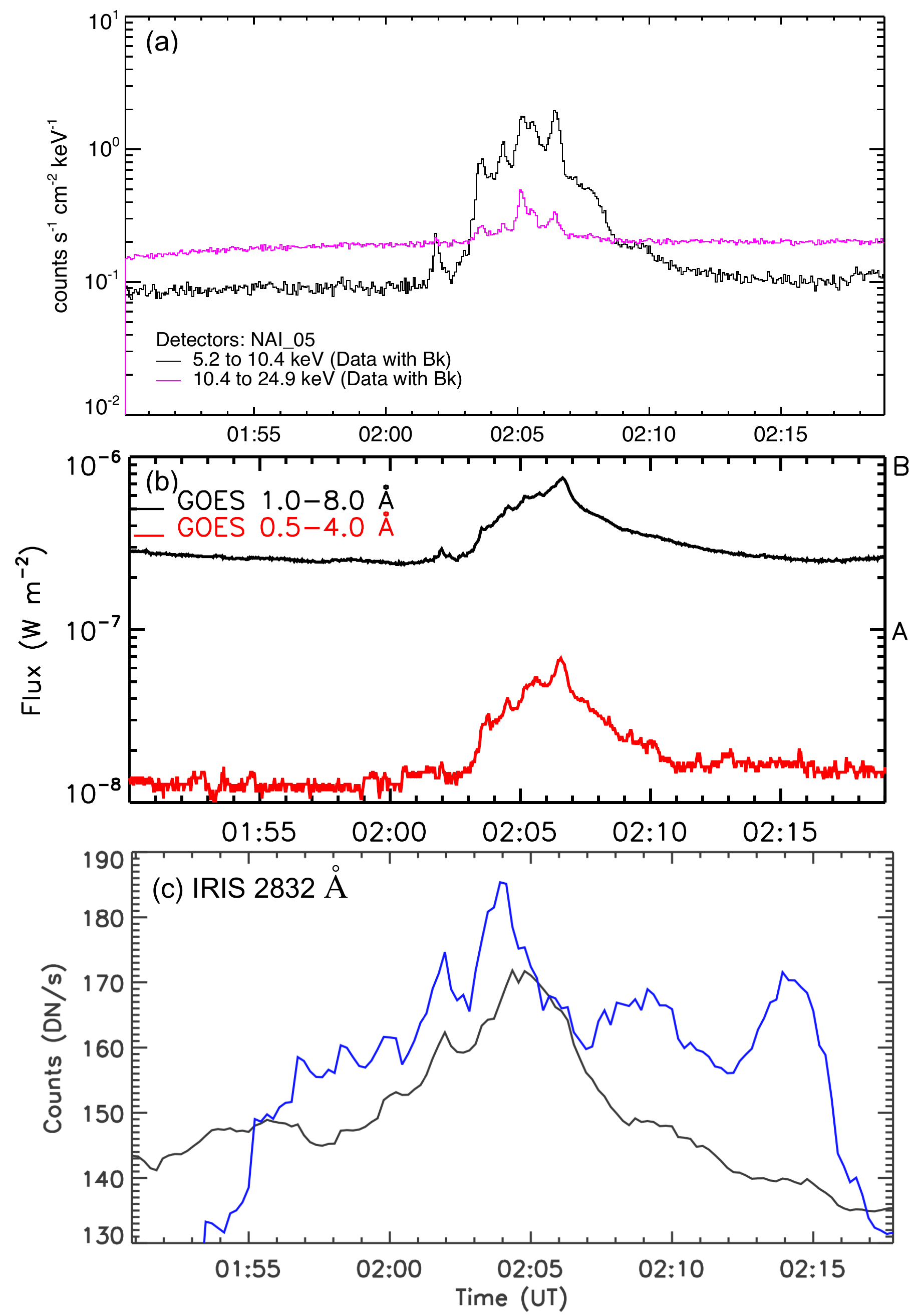}
\caption{
Intensity variation at the flare site observed in FERMI GBM, GOES, and IRIS SJIs. Panel(a): {X-ray count rates detected in two energy channels by FERMI/GBM.}. 
Panel (b): GOES light curve for the B6.7 class solar flare, shows the flare starts at 02:02 UT and peaks at $\approx$ 02:06 UT with small peaks corresponding to  GBM peaks. Panel (c):  {Intensity light curves at the bright points in the IRIS 2832 \AA\  SJIs. The black and blue curves are the light curves over the green and blue boxes presented in Fig. \ref{fig:iris} (d) respectively. 
 The DN signal 
at the flare base is integrated and divided by the total number of pixels.}}
\label{goes}
\end{figure}
\begin{figure}[ht!]
\centering
\includegraphics
[width=0.5\textwidth]
{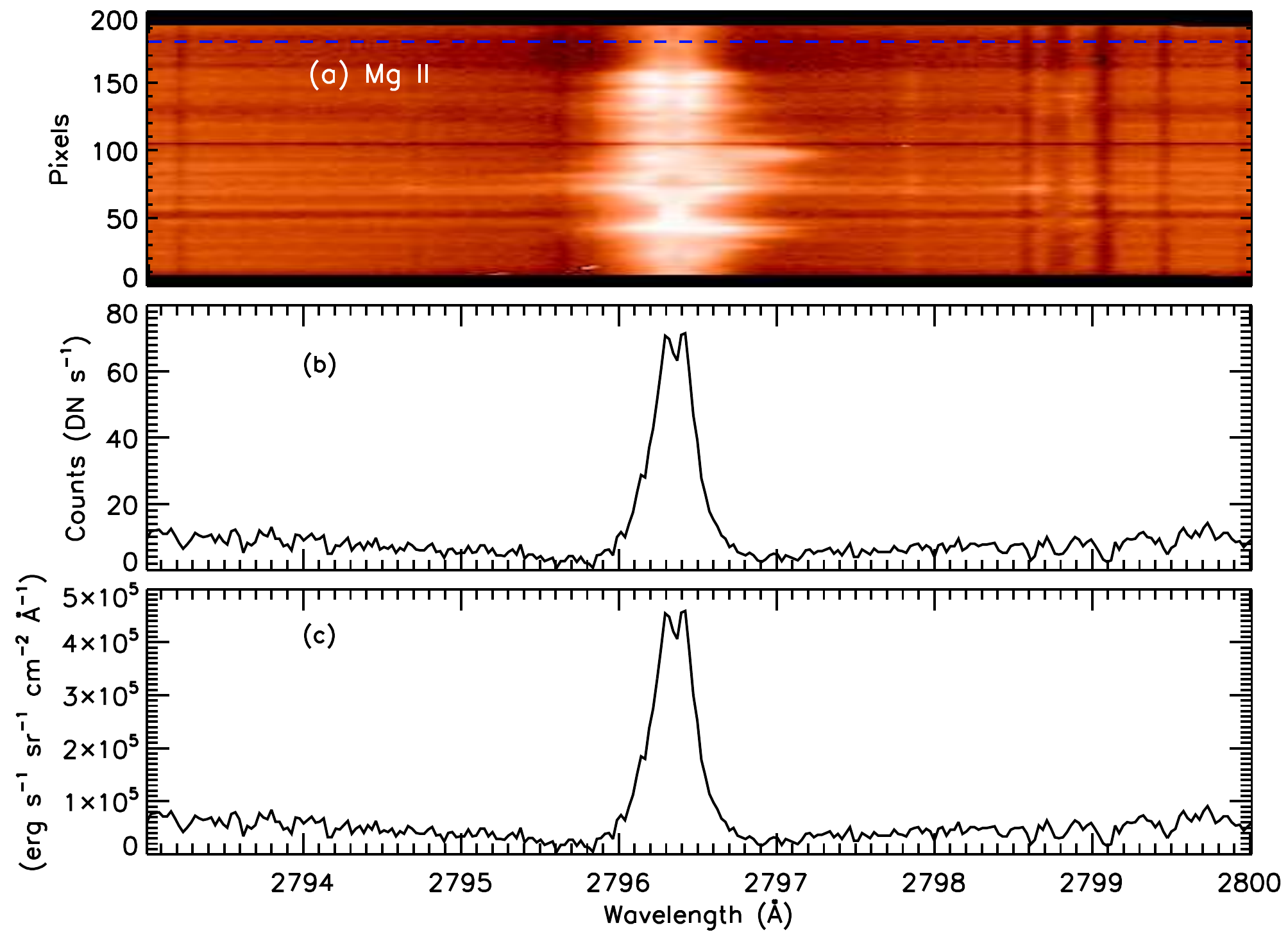}
\caption{Mg II k line spectra along the slit at 01:43:41 UT before the flare (a). Panels (b) and (c) show the Mg II k line profiles in position y = 180 pixel representing the quiet sun in DN s$^{-1}$ and in cgs units, respectively.}
\label{Calib_Nocalib}
\end{figure}
\subsection{Mini flare light curves}
{We first check the occurrence times  of the soft X-ray (SXR) and HXR peaks with the bright point  intensity observed in  the 2832 \AA\ filter by comparing  the time variation of  X-ray flux with
the intensity light curve  of   2832 \AA\ filter between  01:44 UT and  02:20 UT (Fig. \ref{goes}).}
     The SXR light curve of the flare is recorded by  GOES spacecraft in 0.5--4 \AA\ and 1--8 \AA.  
  {The HXR count rates  in different energy bands are recorded} with the Gamma-ray Burst Monitor \citep[GBM,][]{Meegan2009} on board FERMI spacecraft {launched in 2008. FERMI GBM fills the gap for HXR measurements for the solar physics community after the decommissioning of RHESSI.} 
  {FERMI  GBM does not provide HXR images like RHESSI so we cannot check if the HXR emission comes from the flaring active region, but at that time only one active region was present on the solar disk }
 (Fig \ref{fig:fulldisk} left panel) so the HXR emission is expected to come from the flaring active region. 
   In the bottom panel of Fig. \ref{goes}, Mg II 2832 \AA\ slit jaw
    intensity light curves 
    are obtained  by  integrating  the DN signal  divided by the total number of pixels in 
    the small boxes at the flare site (Fig. \ref{fig:iris} green and blue boxes).
We conclude on a good   agreement  between the  occurrence times  of the 
Balmer continuum enhancement, HXR emission  and  the GOES light curves  for this flare.
\begin{figure*}[ht!]
\includegraphics[width=0.9\textwidth]{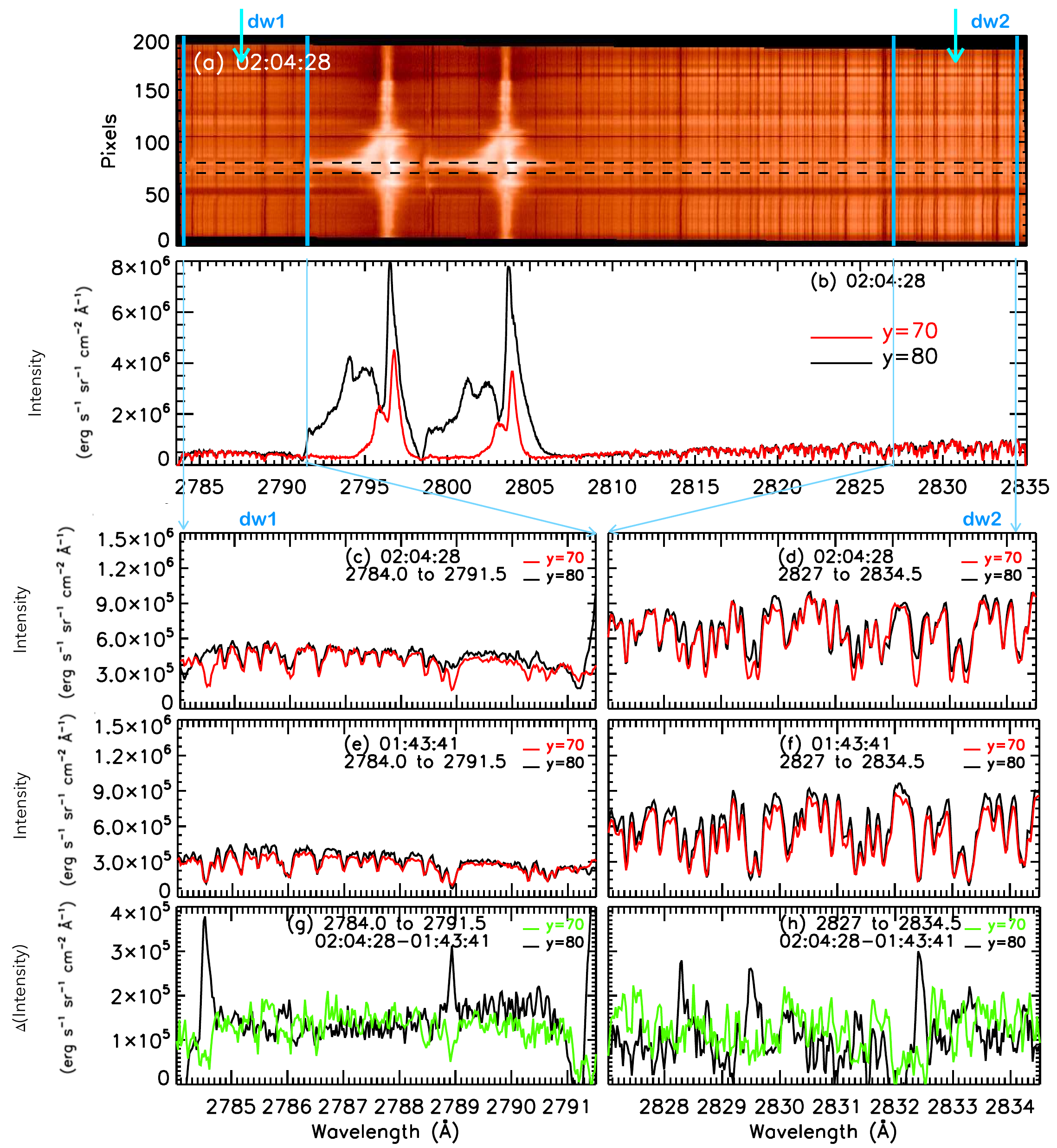}
\caption{Flare intensity distribution in the IRIS Mg II   wavelength  band at the flare time (02:04:28 UT). Panel (a) shows the full spectra of the  Mg II  band   between $\lambda$= 2784 \AA\ and $\lambda$ =2835 \AA\ (50 \AA) along slit 1 position.  Two dashed horizontal black lines indicate the coordinates  y= 70 and y = 80  at the location of the maximum of the Balmer continuum in the spectral image. Panel (b) presents the spectral profile in the full range of the IRIS Mg II wavelength band  at the two  positions (y = 70 and 80 pixels).
Panels (c) and (d) show the  zoom of two  spectral profiles of panel (b) at (y= 70, 80) located at  the two extremities  of the
wavelength  band  (dw1, dw2) indicated  by  two  vertical cyan  lines and   two cyan arrows at the top of panel (a), respectively  in   the right and in the left of the panel. Panel (e) and (f) show the similar  zoom spectra in the two continuum regions at the pre flare time (01:43:41).
Panel (g) and (h) show the difference of intensity  ($\Delta$(Intensity))  between  the intensity at the  flare time (02:04:28 UT)  and at the  preflare time (01:43:41 UT). The intensity is calibrated to the c.g.s. units, {\it i.e.} erg s$^{-1}$ sr$^{-1}$ cm$^{-2}$ \AA$^{-1}$.
}
\label{Calib_all}
\end{figure*}
\begin{figure}
\centering
\includegraphics[width=0.5\textwidth]{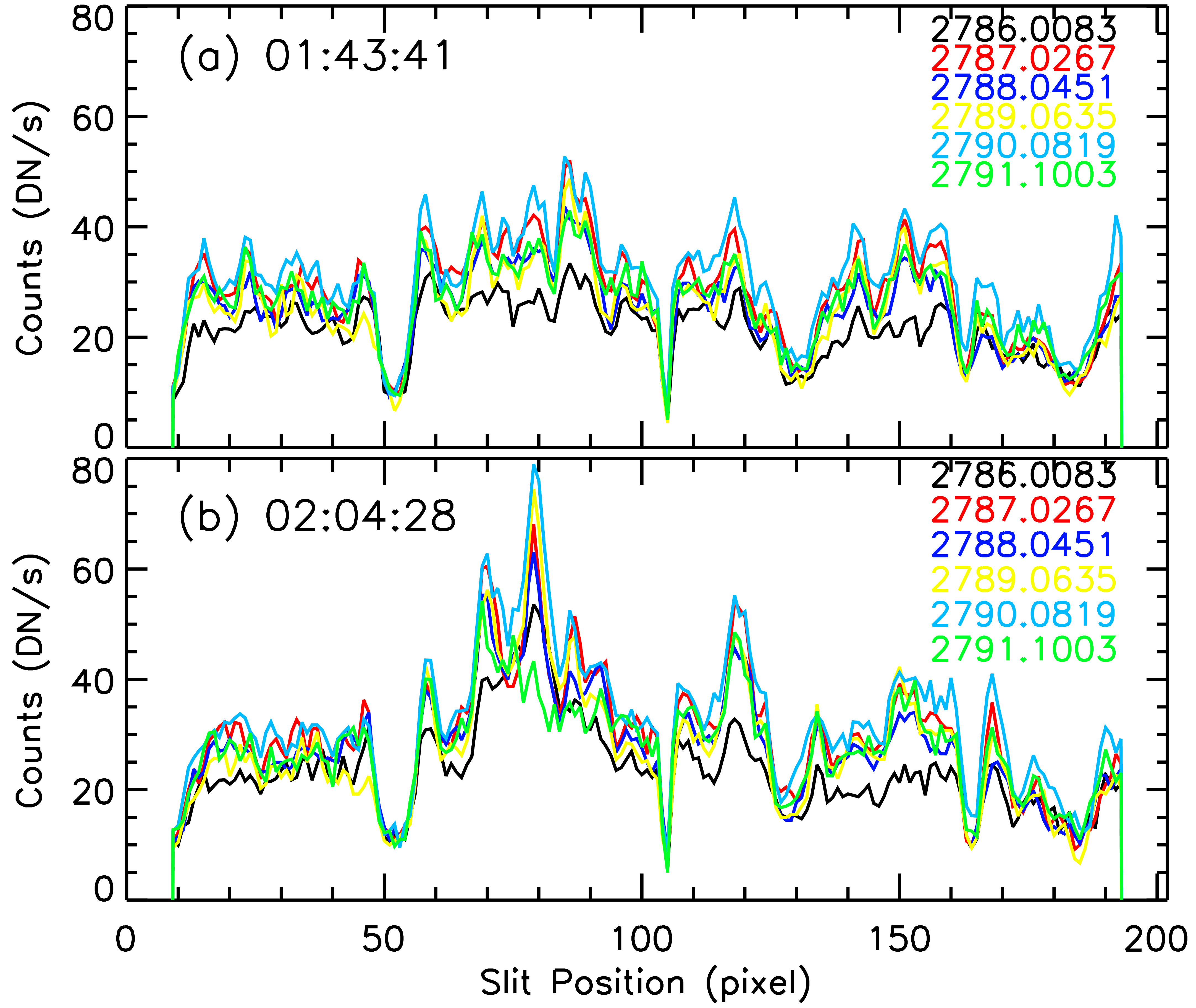}
\caption{Flare intensity variation along the IRIS slit at different wavelengths  {at  the pre-flare (01:43:41 UT) (panel a) and flare (02:04:28 UT) (panel b)  times.
The different colors of the curves are for the   different wavelengths. The intensity variation shows a constancy during the pre-flare time, whereas the curve shows two peaks at y= 70 and 80  pixel during the flare time. {The counts are minimum at the slit position  y= 180  pixel.  These three  positions are  chosen as the reference of the quiet sun and the Balmer enhancement respectively.}}}
\label{balmer_continuum}
\end{figure}

\section{Balmer Continuum enhancement} \label{sec:iris}
{As it is said in the introduction, the enhancement of the Balmer continuum can be interpreted in different ways.
To disentangle between the different mechanisms we need to analyse the IRIS spectra in details (Fig. \ref{balmer_MgII})
and calibrate the   data for 
computing  quantitatively the excess of the Balmer continuum emission.}
\subsection{Intensity calibration spectra}
{Using the IRIS radiometric calibration, we convert 
the data number (DN) to intensity  units.}
We use the IDL routine iris\_calib.pro and get the intensity value (I$_c$) in erg cm$^{-2}$ s$^{-1}$ sr$^{-1}$ \AA$^{-1}$. The exposure time per slit position was $\approx$ 2 sec for this conversion. 
The radiometric intensity calibration is done with a code developed by H. Tian, with the following formula:
\begin{equation}
    I_{c} = \frac{hc}{\lambda} \times n \times \frac{I_{o}}{A_{eff} \times d \times \omega} 
\end{equation}
The photon energy ${hc}/{\lambda}$ is calculated with Planck constant h = 6.63 $\times$ 10$^{-27}$ erg-s, speed of light c = 3 $\times$ 10$^{10}$ cm s$^{-1}$. $n$ is the number of photons per DN, it has a value 4 for FUV and 18 for NUV spectra. I$_o$ is the observed intensity in DN s$^{-1}$ and A$_{eff}$ is the effective area in   cm$^{-2}$ and obtained through $iris\_get\_response$ routine available in SolarSoftWare (SSW). Dispersion $d$ is taken in \AA\ pixel$^{-1}$. The solid angle $\omega$ is calculated as: $\omega = 0.3327 \times 720 \times 0.33 \times 720/({1.5 \times 10^8})^2$.

Figure  \ref{Calib_Nocalib}  shows the {relationship} 
between the DN signal in the range of Mg II k line and the calibrated data for the quiet sun (y = 180 pixel). The Mg II k line peaks at {$4.8 \times 10^5$} erg s$^{-1}$ sr$^{-1}$ cm$^{-2}$ \AA$^{-1}$.  
{In both ends of the domain the photospheric Mg II wing emission is increasing.}

\subsection{IRIS spectra before the flare}
\begin{figure*}
\includegraphics[width=\textwidth]{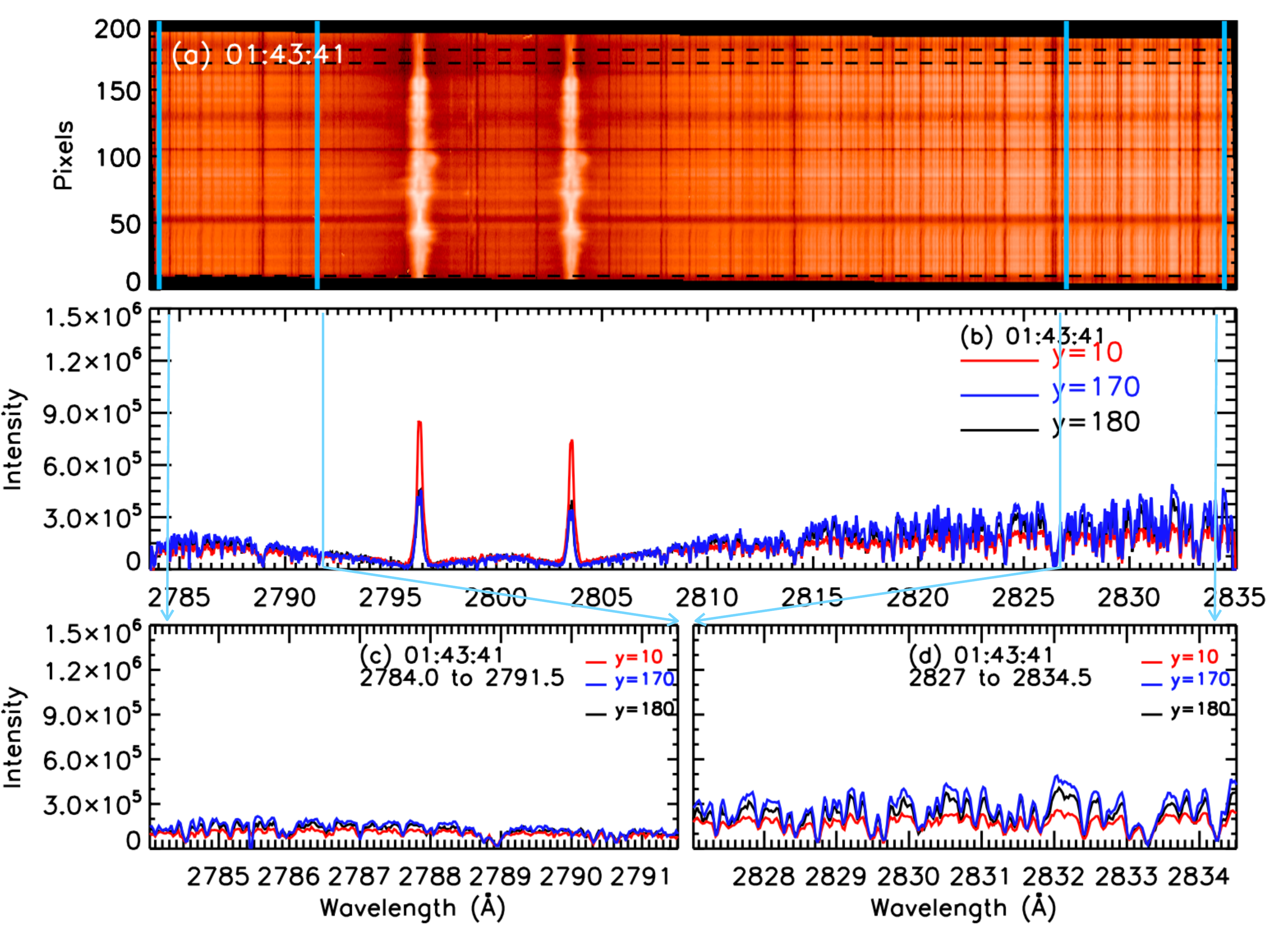}
\caption{Quiet Sun intensity distribution  in the IRIS Mg II   wavelength  band  at the preflare time (01:43:41 UT). Panel (a) shows the full spectra of the  Mg II  band   between $\lambda$= 2784 \AA\ and $\lambda$ =2835 \AA\ (50  \AA), {the dashed  horizontal black lines 
indicate y= 10, 170, and 180 which represent  quiet sun  regions. Panel (b) presents the spectral profile of the full range of Mg II lines for  the three  y pixel values (10, 170, 180). Panels (c) and (d) show the  zoom of two  parts of the  spectral profile 
at the two ends of the wavelength band, shown by 
two cyan vertical lines respectively  in the right and in the left in panel (a) and (b).} The intensity is calibrated 
in erg s$^{-1}$ sr$^{-1}$ cm$^{-2}$ \AA$^{-1}$.}
\label{Calib_QS}
\end{figure*}

 Figure \ref{Calib_QS} panel (a) shows the full range of the Mg II window spectra for the quiet sun before the flare at 01:43:41 UT. This spectra will be our reference spectra.
{The full IRIS wavelength range,  focused on the Mg II lines  and is  extended from 2784 \AA\ to 2835 \AA.
This large domain allows us to analyse the Balmer continuum  relatively far from the cores of the Mg II h and k  lines.  {The spectral profiles of three pixels (y=10, 170, 180) in the quiet sun (Fig. \ref{Calib_QS} panel b) can be compared  to the  global sun spectral profile  shown in  \citet{Pereira2013}.}
Such spectral  profile has a U-shape with an increasing  emission toward the two extremities of the spectral range, in the blue far wing of k and in the far red wing of h, both emissions corresponding to   photosphere emissions (Fig. \ref{Calib_QS} panels  (c) and (d) respectively). In the center of the profile there are the two chromospheric h and k lines  cores in emission. }

Even for the flare time  the  spectral profile
could be about the
same as that detected in the IRIS spectrum outside the flare if there was no Balmer continuum enhancement.
The fact that 
the "continuum" peaks in time
with  the HXR emission  {indicates that there  exists  a  direct relationship between the enhancement of the continuum and the non thermal flux (Fig. \ref{goes}).  However extended  wings of Mg II lines  during flares may also affect the continuum.}
Therefore we explore the wavelength range the farthest as possible from k and h line cores during the flare time.

\subsection{IRIS spectra at the flare time}
{The site of the flare is  crossed  by only one slit of the raster (slit 1) as it was shown in \citet{Joshi2021}. 
Figure  \ref{Calib_all} panel (a) shows the  full  wavelength range of the Mg II spectra along this  slit at the time of the mini flare (02:04:28 UT).
The reconnection  site  is  detected  by {bidirectional outflows}, principally blueshifts  
which concerns    one or two pixels along the slit. 
We note   the  U-shape profile with  an increasing intensity at the two extremities of the wavelength range (Fig.  \ref{Calib_all} panel (b)), similar behaviour of the spectral profile  of  the quiet sun (Fig. \ref{Calib_QS} panel (b)).
 It is difficult to detect the excess of the continuum emission without  proceeding to a subtraction of the quiet sun profile.} First we select the pixels where the Balmer continuum enhancement is the more visible.  
 For that we  made  cuts  in the  Mg II spectra  (Fig.\ref {Calib_all} panel (a)) for different wavelengths in the continuum in the extreme left and  right perpendicularly to the dispersion direction (Fig.  \ref{balmer_continuum}).  Peaks in the continuum at the flare time are well visible for  pixels 70 and 80 along the slit, which we select for the following analysis.

By subtracting the background spectra in the whole IRIS Mg II wavelength band observed before the flare (01:43:41 UT)  from the spectra during the flare  (02:04:28 UT) we 
get the  Balmer-continuum excess. We check that the signal is constant 
at the boundaries of the domain  after  the subtraction. Thus two different domains of  the continuum  are chosen, {\it e.g.} 2784-2791.5 \AA\ and 2827-2834.5 \AA\  to compute   the intensity variation (Fig.  \ref{Calib_all} panels c-d for flare time, in panels e-f for pre-flare time) and finally the subtraction of both  intensities 
(Fig. \ref{Calib_all} panels g-h).
The difference intensity plots show a constant continuum enhancement  in each selected domain of the order of 1.5 to 1.75 $\times$ 10$^5$ erg s$^{-1}$ sr$^{-1}$ cm$^{-2}$ \AA$^{-1}$. The  superimposed  residual signal is weak, due to optically-thin Balmer emission but significant.
{The  enhancement of the Balmer continuum corresponds to an increase of {about 50 $\%$} over the pre-flare level during the flare}.

\begin{figure}[ht!]
\centering
\includegraphics[width=0.5\textwidth]{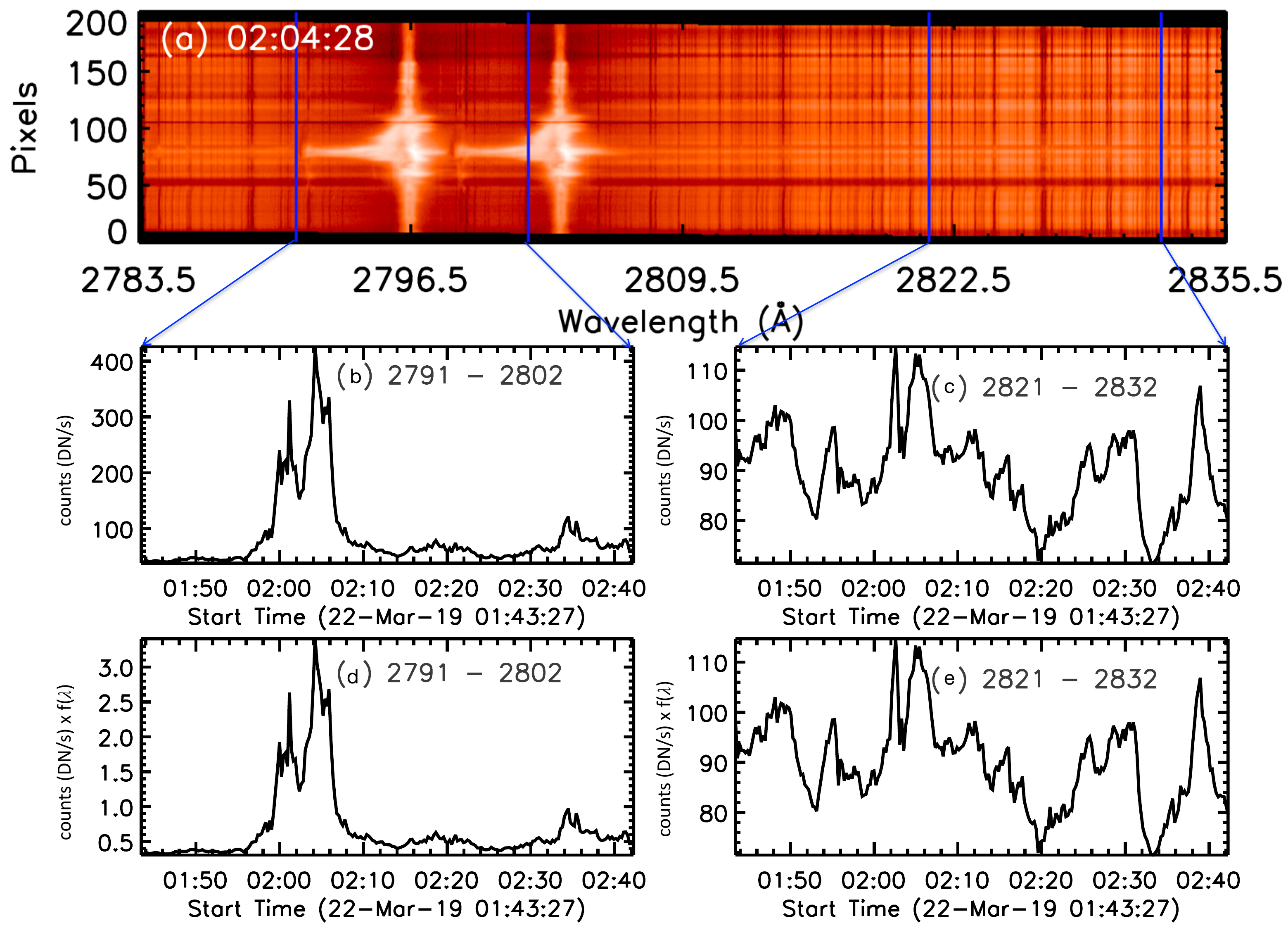}
\caption{Evolution versus time of the intensity in the Mg II k  and in the 2832 \AA\ ranges. Top panel: IRIS spectra of the   Mg II band along the slit position 1. The two wavelength regions in Mg II k line (left) and in the Balmer emission (right) are bounded by vertical blue lines.
Bottom panels (b-e): temporal evolution of the relative contribution of Mg II k line (b, d)  and Balmer continuum  emission  (c, e) at the reconnection point of the flare (y=80).  In panels c and d the intensity (DN) has been divided by the transmission  factor f($\lambda$) of the  2832 filter taken from \citet{Kleint2017}.
The comparison of the  relative  peak values 
at the flare time  in Mg II range  (panel c) and continuum region (panel d) indicates the prevalence
of the Balmer continuum.}
\label{sji2832}
\end{figure}

\subsection{IRIS SJI 2832 \AA\ filter}
 The SJIs in the 2832 \AA\   filter  show  a brightening at the location of the flare around 02:05 UT  (Fig. \ref{fig:iris}). In Fig. \ref{goes} we estimate the increase of the brightening  in the small box (Fig. \ref{fig:iris} panel(c)) by 50 $\%$ ((180-120)/120) between the time of the pre-flare (01:44 UT) and the flare time (02:04 UT). {The SJI 2832 \AA\  filter includes the  two 
 Mg II k and h  lines and a  large wavelength domain of the continuum. {The 2832 filter transmission profile  f($\lambda$)  has two peaks: one peak  around  2830  \AA\ and the second peak close to the k line \citep{Kleint2017}).}  
 The ratio  of the transmission profile between the two peaks is 10$^{-2}$. 
 The 2832 \AA\ filter reduces the intensity of Mg II h  and k lines by a factor nearly equal to 
 10$^{-2}$  compared to 
 the continuum emission} (see Fig. 3 in \citet{Kleint2017}). 
 
 {With the presence of the two peaks of f($\lambda$) the  increase of the brightening in the SJI 2832 \AA\ 
 may be contaminated by h and k lines and Fe lines present in   the  two wavelength domains: h and k and near 2830 \AA\ \citep{Kleint2017}.
 By chance we are able to   eliminate this scenario 
 because  we have the Mg II spectra at the same time which allows us  
 to understand  what is the contribution of the lines versus  the continuum.  We consider again the  Mg II spectra obtained simultaneously in the  IRIS rasters (slit 1) and in the SJIs   before and during the flare time.}
 So that  we  consider two wavelength ranges,  one range containing the Mg II k and h lines and one range around 2832 \AA. The  two light {curves} versus time  for each wavelength range show a  peak at the time of the flare (Fig. \ref{sji2832} panels b and c).

During the flare at 02:04 UT  the  peak maximum of the DN/s integrated over the wavelength range 2791 -2802  \AA\  (h and k range) is  equal to 3 and 27 (112-85) for the continuum (2821-2832 \AA) after applying the reduction factor of the filter. The excess of the Balmer continuum is around 30$\%$  (Fig. \ref{sji2832}  panels d, e).  It is less that what we  compute directly with the spectra  and the SJI  but the wavelength ranges are wider including the two Mg II lines and for the continuum  domain it  is noisy due to the presence of Fe {lines}.  

  Further on we check the contribution of the h and k lines  in the excess of Balmer continuum   with  their calibrated values after applying the decrease factor due to the 2832 \AA\ filter.   
 In Fig. \ref{filter}, we present the calibrated intensity difference  curves in three wavelengths domains between the pre-flare  and flare time, two  curves correspond to the continuum (Fig.\ref{filter} panels a, c) and one curve the Mg II h and k line (Fig. \ref{filter} panel b). After  multiplying by the transmission profile  of the filter we obtain the values of the intensity in the line (peak at 6 $\times$ 10$^4$ erg s$^{-1}$ sr$^{-1}$ cm$^{-2}$ \AA$^{-1}$) and in the continuum (1 $\times$  10$^3$ erg s$^{-1}$ sr$^{-1}$ cm$^{-2}$ \AA$^{-1}$) in the band around 2787 \AA\ and {(1.5-1.75 $\times$ 10$^5$ erg s$^{-1}$ sr$^{-1}$ cm$^{-2}$ \AA$^{-1}$) } in the band around 2832 \AA\   ).
  The emission of the k line and at 2787 \AA\  is very much reduced  due to the reduction factor of  the filter.
  
  In a second step we compute the integrated value of the k line over the  wavelength  range (2790-2802 \AA) and compare to the integrated intensity of the continuum (2820- 2832 \AA) (Fig. \ref{filter}  panels b and c).  
 {The ratio between the total shaded areas in panel b (20110 for 472 pixels) and in panel c (101348 for 589 pixels) gives the contribution of the Mg II lines (around 18\%) compared to the Balmer continuum (82\%) in the wavelength domain of the 2832 filter  during the flare.}

  In conclusion, the main contribution (82$\% $) of the emission in the 2832 SJIs is the Balmer continuum around 2832 \AA\ during our mini flare.
  It confirms the results of 
 \cite{Kleint2017} showing that the Balmer continuum  enhancement could be  the principal  contributor to the   flare brightenings  in the 2832 \AA\ SJIs  in very specific pixels. 
 In fact their observations concerned  an  X-class   solar flare  where the Balmer continuum enhancement can also be affected by  the Fe II line emission. As our flare is very weak (B-class) no Fe II emission is  detected. It is the reason why our conclusion is more convincing about the enhancement of the  Balmer continuum.

\section{FERMI measurements of HXR emission }
\label{sec:fermi}
\subsection{FERMI spectral analysis}
{A significant proportion of energy released during a solar flare is thought to go into particle acceleration \citep{Emslie2012}. By studying the electron spectra obtained from the HXR photons during the flare phase one can deduce important information about the energetics of the non-thermal electrons.}
{For our study, we use FERMI/GBM data (12 NaI detectors) which cover the range between 8 keV to 1 MeV and provides  128 quasi-logarithmically spaced energy bins with 4.096 s temporal resolution for several  detectors with different viewing angles and we select data from the most sunward NaI detector (detector 5).}

{Figure \ref{goes} shows the FERMI/GBM count-rates in 2 energy bands for the most sunward NaI detector (detector 5) for the time period of the IRIS event analysed in the previous section. Several HXR peaks are detected by FERMI/GBM below 14.6 keV at the time of the GOES B-class flare with one of these peak detected above 14.6 keV (time interval 02 :04 :43 to 02 :06 :34 UT). The spectral analysis performed during  this time interval  between 8 and 25 keV is shown on Fig. \ref{fig:fermi3}.}
Using the OSPEX module in SSW, we fit a variable {isothermal model} (blue line in the Fig. \ref{fig:fermi3}) and the non-thermal thick target component {\it thick2} which directly provides the non-thermal electron spectrum {from the fit} (red line in the Fig. \ref{fig:fermi3}). 
The electron spectrum (Fig. \ref{fig:fermi3}) is chosen as a broken power-law. The best fit is obtained for the following parameters: spectral index -2.97 between 13.9 keV and 62.7 keV and -5.2 above 62.7 keV. The number of non-thermal electrons produced in the flare as well as the non-thermal energy contained in these non-thermal electrons can be deduced from these parameters. One of the main uncertainty on these numbers arises from the determination of the low energy cutoff because of the dominant thermal emission at low X-ray energies.

The best fit is obtained for a low energy cut-off of 13.9 keV but allowing a variation of the minimum chi-square of 2\%  gives a range of possible values for the low energy cut-off going from 10.3 to 19.5 keV. Computing the number of energetic electrons above the cut-off gives number in the range: 1.51 $\times$ 10$^{33}$- 4.86 $\times$ 10$^{32}$. 
For comparison with the production of the Balmer continuum, a useful quantity is the non-thermal energy input rate per unit area. Since FERMI does not have imaging capability we are unable to measure the area of the HXR emitting region, but we can expect from HXR measurements with RHESSI that the HXR footpoint area is of the order of 3\arcsec ({\it e.g.} \citet{Dennis2009} for HXR footpoint measurements). {An area  of 
    3 arcsec$^2$ (3\arcsec x 1\arcsec) 
    would lead to an energy input
 rate of 5.4 $\times$ 10$^{8}$ erg s$^{-1}$ cm$^{-2}$.} Another estimate for the area can be made using IRIS observations, as it was suggested in \citet{Kleint2016} where the area was evaluated by measuring the cross section of the ribbon along IRIS slit. The spatial resolution of IRIS is 0.33$\arcsec$.

{Thus if we assume an upper-limit of 0.5 $\times$ 0.5 arcsec for the area of the deposition of non-thermal energy,  the energy input rate per unit area produced by electrons above 20 keV will be {6.5 } $\times$ 10$^{9}$ erg s$^{-1}$ cm$^{-2}$.}

\begin{figure}[t!]
\centering
\includegraphics[width=0.49\textwidth]{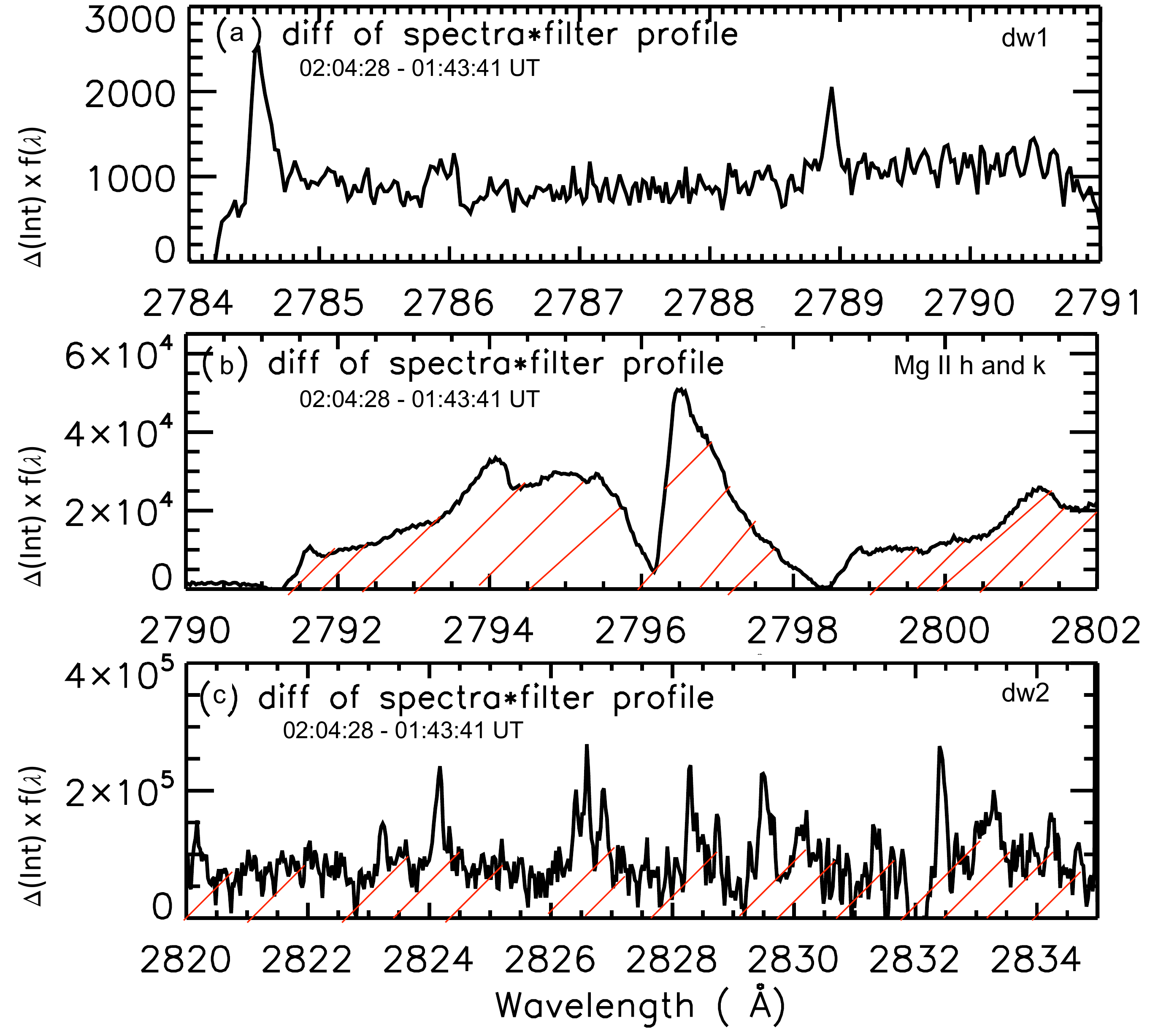}
\caption{Variation with  the wavelength  of the difference of intensity between flare and preflare times in three wavelength ranges containing continuum regions (dw1 and dw2) and Mg II h $\&$ k line. 
The differences of intensity have been multiplied by the 
transmission filter profile f($\lambda$) of  the 2832 \AA\ filter (\citep{Kleint2017}. After normalization of the hatched areas in panels (b and c), we estimate that the contribution of the  excess of the Balmer emission   is 82 \% compared to the contribution of the  Mg II k line around  18$\%$  in the wavelength domain of the 2832 \AA\  SJI during the flare.}
\label{filter}
\end{figure}

\begin{figure*}[t!]
\centering
\includegraphics[width=\textwidth]{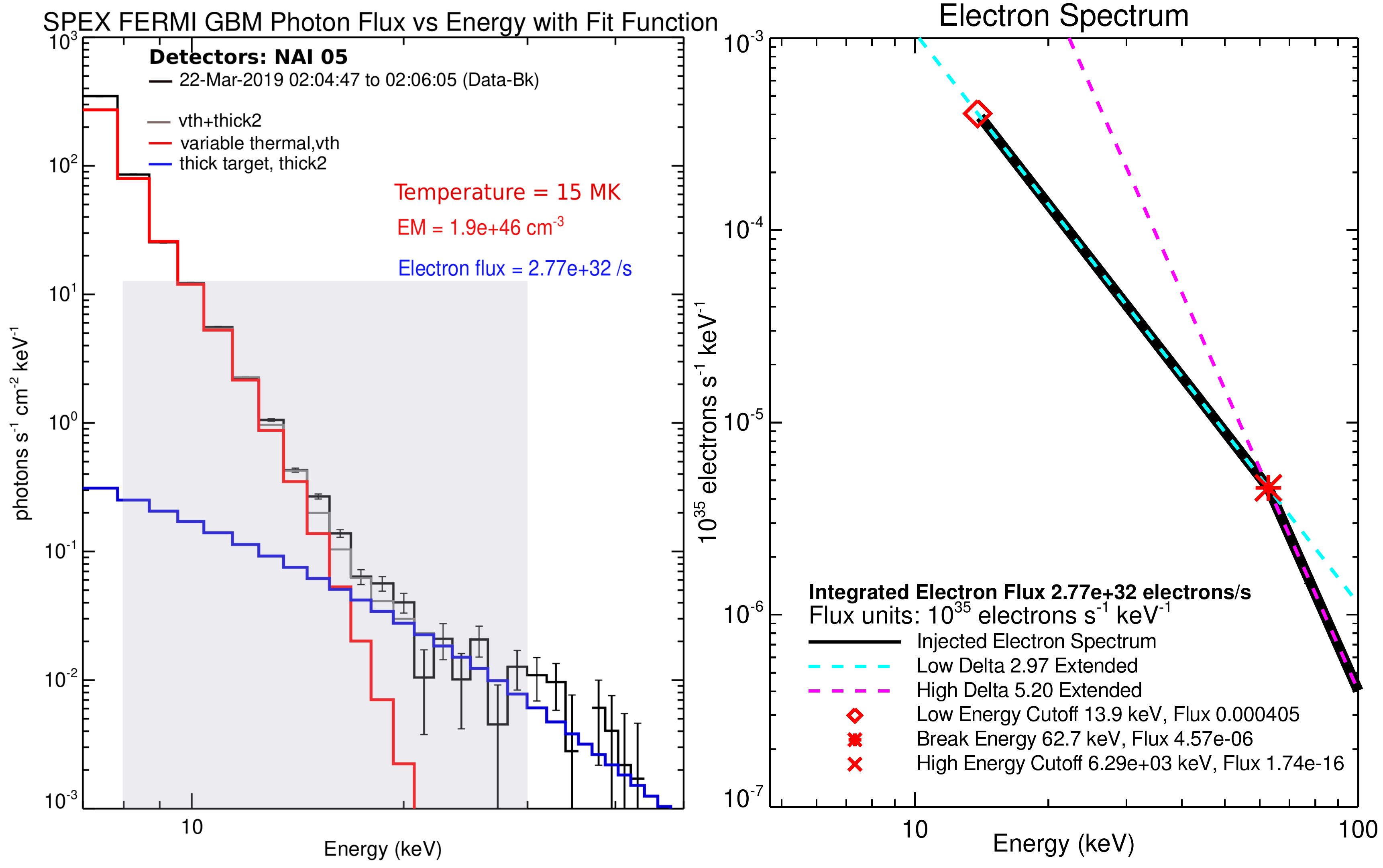}
\caption{
FERMI GBM photon spectra accumulated around 02:04:47 UT and 02:06:05 UT during the quoted 2-minute interval on March 22  2019 (left panel). 
{Variable {isothermal} and thick target fitting done to the FERMI GBM spectrum. For variable thermal we use `vth' function and for thick target we use `thick2' from OSPEX. The variable thermal fit is shown by the red curve and the thick target fit is shown by the blue curve. The fit is limited between 8-30 keV shown by the shaded region. Above 30 keV counts are predominantly that of background. The fit shown here is best fit obtained using $\chi^2$ as a measure. We find a range of values for low-energy cutoff parameter (10.3 - 19.5 keV) gives reasonable fit.}
Electron spectrum given by the thick target fit is shown on the right panel.}
\label{fig:fermi3}
\end{figure*}

\subsection{Non-LTE radiative transfer models}
The second step of our study was to estimate  if the energy needed for  the Balmer continuum excess  could correspond to the energy brought  by the bombardment of non thermal electrons. We need so to  compare the energy {input }provided by non thermal electrons  during the mini flare with the excess of the Balmer continuum enhancement observed by IRIS.  This kind of comparison was done previously in the paper of   
\citet{Kleint2016}   where they presented  IRIS and RHESSI data for a strong  X-ray flare. 
For that the former authors considered {a theoretical grid  of 1D static flare atmosphere models developed by \citet{Ricchiazzi1983},  usually referred as RC models, in which the temperature structure is computed via the energy-balance  between the electron-beam heating, conduction and net radiation losses and using as a free parameter the enhancement of the coronal pressure due to evaporation.  Further, \citet{Kleint2016} used the non-LTE code MALI to synthesize the hydrogen  recombination continua for all the models. They compared the resulting  intensities of  the Balmer continuum  to those detected by IRIS.}
The RC flare models thus provide a relationship between the electron-beam energy flux with the cut-off energy 20 keV
having a given spectral index and the enhancement of the Balmer continuum. Our observed excess of the Balmer continuum intensity is of the order of 
1.5 to 1.75 $\times$ 10$^5$ erg s$^{-1}$ sr$^{-1}$ cm$^{-2}$ \AA$^{-1}$  (Fig. \ref{Calib_all} panels g and h, Fig. \ref{filter} panel c).
According to results of \citet{Kleint2016} (see their Table 1) this value is roughly consistent with a beam flux of  10$^9$ and  10$^{10}$ {
erg s$^{-1}$ cm$^{-2}$}
(RC models E5 and E12). Such fluxes have been derived from FERMI GBM observations as we show above. For model E5 the coronal pressure is 100 dyn cm$^{-2}$  which leads to very bright Mg II line cores as computed in \citet{Liu2015} which are not observed in our weak flare (such a high pressure is found in an X-class flare atmosphere analyzed by \citet{Liu2015}. The second model E12 with lower coronal pressure predicts the Mg II line-core intensities more compatible with our IRIS observations and is thus reasonable for this weak flare. 
The energy value  is of same order than the energy brought by the electron beams of FERMI GBM between {10$^{9}$ and 10$^{10}$ 
{erg s$^{-1}$ cm$^{-2}$} due to  the uncertainty of the area size}.

\section{Discussion and conclusion}
\label{dis}
We extend the analysis of the IRIS data of GOES B6.7 micro flare (or mini flare) observed on March 22, 2019 focusing on  the IRIS SJIs in the 2832 \AA\ filter and the NUV continuum  spectral variation in the two extreme wavelength domains  of the  IRIS 2896 \AA\ spectra bandpass (between 2784-2791.5 \AA\ and 2827-2834.5 \AA).  The mini flare  and its site of reconnection identified by {bidirectional outflows} concern only  one or two pixels along  slit position 1 of the IRIS rasters
\citep{Joshi2021}. Therefore we focus our analysis on the spectra along this slit.
Through the paper we demonstrate that the enhancement of the NUV continuum was due to electron beams and not to the flare  heating by  inter-playing   the analysis of  IRIS images  and spectra.

A brightening in the 2832 filter is visible  in the  area of the flare.
After considering  the wavelength domain of the filter, its transmission function, and the emission of  the spectra in the same spectral domain we conclude that  it was the direct signature of electron beams. In fact the  excess of the Balmer continuum in the spectra detected at the flare time in the  few  considered pixels is constant all along the two considered spectral ranges
(bottom  panels of Fig. \ref{Calib_all}). 
The increment in  the continuum  intensity being constant can not be due to the intensity  enhancement in the far wings of Mg II lines, which is 
decreasing  with distance from the k (2796 \AA) or h (2803 \AA) Mg II line center on either side. Balmer continuum  being  optically-thin in a weak flare, thus its emission  really   far from Mg II h (more than 30 \AA)  for the second range is simply added to the photospheric (pre-flare) background as {a constant excess of intensity} over the photospheric continuum.

The co-temporal time  of the continuum enhancement peak in the  spectral profiles and in  SJIs suggests that the SJI brigthness is dominated by the Balmer enhancement continuum.
We found  that the contribution of the Balmer continuum  is of the order of  82 \% compared to the 18 \% of the extended  Mg II line wings  in the  spectra domain of the 2832 filter. 
Besides the NUV continuum in this wavelength range is relatively simple to analyse, because the metallic lines  (mainly Fe II lines)  present in this domain are not in emission for this mini  flare. This is a very different case from  previous studies concerning X-class flares  where   metallic lines in the spectra  around 2830 \AA\  went into emission  and affected the Balmer continuum  \citep{Heinzel2014,Kleint2017}. The previous authors had to disentangle between the two emissions: continuum and Fe II lines.

For our weak flare we demonstrate that the FERMI GBM energy output by non thermal electrons is  consistent with   the beam flux required  in   non-LTE radiative models  for getting  the  Balmer continuum emission excess  measured in the IRIS spectra. In large flares, it has been shown that the injected non thermal energy  {derived from} RHESSI data is sufficient and even in excess  for 
explaining the thermal component of  strong flares
\citep{Aschwanden2017,Kleint2016}. However it was shown that for weak flares there was a deficit of energetic electrons for affecting the low levels of the atmosphere   where the bolometric emission (NUV, White-light, near IR radiation) is initiated \citep{Warmuth2020}.
\citet{Inglis2014, Warmuth2020}  notice  an apparent deficit of non thermal electrons in weak flares by computing the ratio between thermal
energy (and losses) and energy in non thermal electrons.
The interpretation depends on where  and in which area the electrons are accelerated. In  the strong energetic flares  studied by \citet{Heinzel2014}, and \citet{Kleint2017} the electron beam  energy was large enough to power the thermal flare as the \citet{Ricchiazzi1983} models demonstrated.

In our case the reconnection site of the mini flare at the base of the jet has been identified  in a tiny bald patch region transformed dynamically in a  X-point  current sheet which explains its multi-thermal components \citep{Joshi2020FR}. The electron beam input should be sufficient to power the thermal flare observed with  Balmer continuum excess. The estimation of the non thermal energy is based on the size of the deposit electron area. This is a relatively unknown variable. The site of reconnection may be smaller than the IRIS spatial resolution and the energy input per unit area  is may be underestimated. The spectral signatures of the mini flare have been identified as IRIS bomb  spectra \citep{Peter2014,Grubecka2016,Young2018,Joshi2021}. Such structures can be due to plasmoid instablity which creates tiny multi-thermal plasmoids not resolved by our telescopes \citep{Ni2016,Baty2019,Ni2021}.

This study shows that the {non-thermal} HXR emission detected by FERMI GBM is strongly related to the enhancement of the Balmer continuum emission, signature of a significant  excess of heating, therefore we conclude that the detection of the Balmer continuum emission in our mini flare supports the scenario of hydrogen recombination in flares after a sudden ionization at chromospheric layers.

\begin{acknowledgements}
{We thank the referee for his/her numerous insightful comments which have greatly helped to improve the manuscript.} IRIS is a NASA small explorer mission developed and operated by LMSAL with mission operations executed at NASA Ames Research center and major contributions to downlink communications funded by ESA and the Norwegian Space Centre. We thank Jana Kasparova for the discussion and suggestions.
We are thankful to Hui Tian and Krzysztof Barczynski for the discussion regarding the IRIS data calibration.
RJ thanks to CEFIPRA for a Raman Charpak Fellowship (RCF-IN-00136) under which this work is initiated at the Observatoire de Paris, Meudon. RJ also acknowledges the support from Department of Science and Technology (DST), New Delhi, India as an INSPIRE fellow. PH acknowledges support from the Czech Funding Agency, grant 19-09489S. RC acknowledges the support from Bulgarian Science Fund under Indo-Bulgarian bilateral project, DST/INT/BLR/P-11/2019. 
\end{acknowledgements}

\end{document}